\documentclass[pdflatex,sn-nature,Numbered]{sn-jnl}

\usepackage{graphicx}%
\usepackage{multirow}%
\usepackage{amsmath,amssymb,amsfonts}%
\usepackage{amsthm}%
\usepackage{mathrsfs}%
\usepackage[title]{appendix}%
\usepackage{xcolor}%
\usepackage{textcomp}%
\usepackage{manyfoot}%
\usepackage{booktabs}%
\usepackage{algorithm}%
\usepackage{algorithmicx}%
\usepackage{algpseudocode}%
\usepackage{listings}%
\usepackage{siunitx}%
\usepackage{orcidlink}

\begin{document}

\title[Article Title]{The Simons Observatory: Production-level Fabrication of the Mid- and Ultra-High-Frequency Wafers}


\author*[1]{\fnm{Shannon M.} \sur{Duff} \orcidlink{0000-0002-9693-4478}} \email{shannon.duff@nist.gov}

\author[1]{\fnm{Jason} \sur{Austermann} \orcidlink{0000-0002-6338-0069}}

\author[1]{\fnm{James A.} \sur{Beall} \orcidlink{0000-0003-1263-6738}}

\author[1,2]{\fnm{David P.} \sur{Daniel} \orcidlink{0009-0009-9335-2534}}

\author[1]{\fnm{Johannes} \sur{Hubmayr} \orcidlink{0000-0002-2781-9302}}

\author[1,3]{\fnm{Greg C.} \sur{Jaehnig} \orcidlink{0000-0001-8697-0064}}

\author[4]{\fnm{Bradley R.} \sur{Johnson} \orcidlink{0000-0002-6898-8938}}

\author[1,3]{\fnm{Dante} \sur{Jones} \orcidlink{0009-0005-1195-2458}}

\author[1]{\fnm{Michael J.} \sur{Link} \orcidlink{0000-0003-2381-1378}}

\author[1]{\fnm{Tammy J.} \sur{Lucas} \orcidlink{0000-0001-7694-1999}}

\author[5]{\fnm{Rita F.} \sur{Sonka} \orcidlink{0000-0002-1187-9781}}

\author[5]{\fnm{Suzanne T.} \sur{Staggs} \orcidlink{0000-0002-7020-7301}}

\author[1,3]{\fnm{Joel} \sur{Ullom} \orcidlink{0000-0003-2486-4025}}

\author[5]{\fnm{Yuhan} \sur{Wang} \orcidlink{0000-0002-8710-0914}}

\affil*[1]{\orgdiv{Quantum Sensors Division}, \orgname{National Institute of Standards and Technology}, \orgaddress{\street{325 Broadway}, \city{Boulder}, \postcode{80305}, \state{CO}, \country{USA}}}

\affil[2]{\orgdiv{Simons Foundation}, \orgaddress{\street{160 5th Avenue}, \city{New York City}, \postcode{10010}, \state{NY}, \country{USA}}}

\affil[3]{\orgname{University of Colorado Boulder}, \orgaddress{\city{Boulder}, \postcode{80309}, \state{CO}, \country{USA}}}

\affil[4]{\orgdiv{Department of Astronomy}, \orgname{University of Virginia}, \orgaddress{\city{Charlottesville}, \postcode{22904}, \state{VA}, \country{USA}}}

\affil[5]{\orgdiv{Department of Physics}, \orgname{Princeton University}, \orgaddress{\city{Princeton}, \postcode{08540}, \state{NJ}, \country{USA}}}

\abstract{The Simons Observatory (SO) is a cosmic microwave background instrumentation suite in the Atacama Desert of Chile. More than 65,000 polarization-sensitive transition-edge sensor (TES) bolometers will be fielded in the frequency range spanning 27 to 280~GHz, with three separate dichroic designs. The mid-frequency 90/150~GHz and ultra-high-frequency 220/280~GHz detector arrays, fabricated at NIST, account for 39 of 49 total detector modules and implement the feedhorn-fed orthomode transducer (OMT)-coupled TES bolometer architecture. A robust production-level fabrication framework for these detector
arrays and the monolithic DC/RF routing wafers has been developed, which includes single device prototyping, process monitoring techniques, in-process metrology, and cryogenic measurements of critical film properties. Application of this framework 
has resulted in timely delivery of nearly 100 total superconducting focal plane components to SO with $88\%$ of detector wafers 
meeting nominal criteria for integration into a detector module: a channel yield $>95\%$ and $T_{\mathrm{c}}$ in the targeted range.}


\maketitle

\section{Introduction}\label{intro}
The Simons Observatory (SO) is a cosmic microwave background (CMB) instrumentation suite situated at an elevation of 5200~m in the Atacama Desert of Chile \cite{science-so,galitzki-so}. The baseline observatory consists of one six-meter large-aperture telescope to characterize small angular scale celestial features and three half-meter small-aperture telescopes to search for gravitational waves at large angular scales to probe inflation. SO will field more than 65,000 polarization-sensitive transition-edge sensor (TES) bolometers spanning 27 to 280~GHz, with three separate dichroic designs: low-frequency (LF), mid-frequency (MF), and ultra-high-frequency (UHF) with band centers at 27/39~GHz, 90/150~GHz, and 220/280~GHz, respectively. The LF detector arrays utilize the lenslet-coupled sinuous antenna architecture and are fabricated at UC-Berkeley \cite{obrient-sinuous3,mangu_solf}. The MF and UHF detector arrays, which implement the feedhorn-fed orthomode transducer-coupled architecture \cite{truce1,truce2}, and the DC/RF routing wafers, which provide the cold readout interface between MF/UHF detectors and microwave multiplexers, are fabricated at NIST and account for 39 of 49 total detector modules \cite{healy_umm,mccarrick_ufm,mccarrick_mfufm,healy_uhfufm}.

The evolving science goals of CMB experiments demand larger numbers of densely-packed large-format focal planes of high-yield multi-chroic detectors and readout. To meet this need, the fabrication team at NIST has developed a robust fabrication framework which allows for reliable high-throughput production of science-grade MF and UHF detector arrays and monolithic DC/RF routing wafers. Implementation of this framework has resulted in delivery of $>$40 150-mm high-yield superconducting TES bolometer arrays and $>$50 DC/RF routing wafers; the in-lab cryogenic detector array performance is described in a companion paper \cite{dutcher_somf}. 
We present the fabrication process (Sec.~\ref{sec:fab}), including in-process and post-process metrology (Sec.~\ref{sec:metrology}), and discuss best practices which allow for higher volume production in a shared R\&D cleanroom (Sec.~\ref{sec:production}). Section~\ref{sec:dualtes} summarizes recent work aimed to facilitate in-lab characterization of TES bolometer arrays via the addition of a second in-series TES.

\section{Fabrication Process Overview}\label{sec:fab}

\begin{figure}[h]%
\centering
\includegraphics[width=0.99\textwidth]{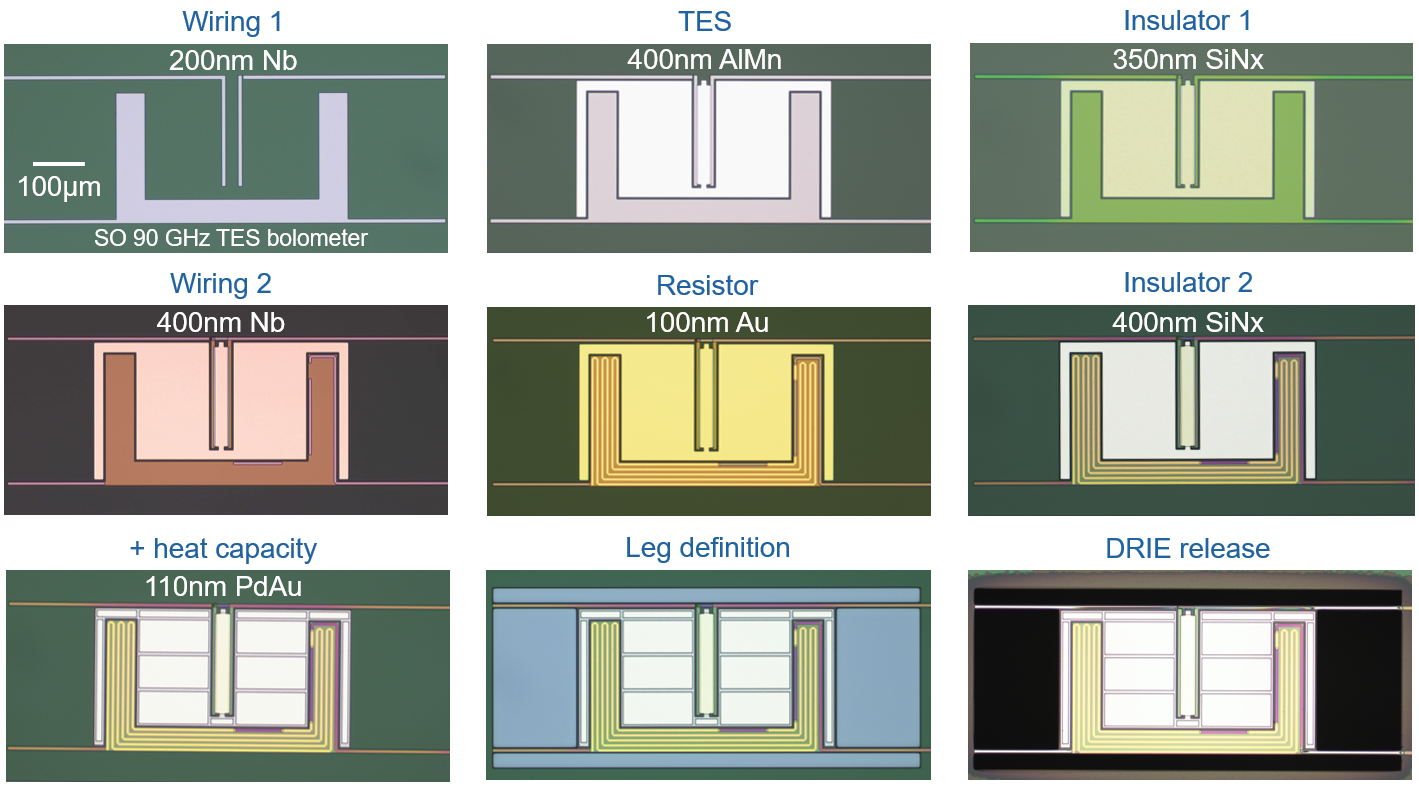}
\caption{Layer-by-layer micrographs of a 90~GHz TES bolometer from a SO MF detector array.}\label{fig:fab}
\end{figure}

\begin{table}[b]%
\caption{Detector array and DC/RF routing wafer fabrication steps.}\label{tab:fab}
\begin{tabular}{@{}llll@{}}
\toprule
Layer/Material                                 & Thickness (nm)           & Deposition Process & Etch Process             \\
\midrule
Thermal SiO$_x$\textsuperscript{1,2}               & 450\footnotemark[1], 150\footnotemark[2] & Thermal            & CF$_4$/Ar/O$_2$ ICP-RIE  \\
Super-low-stress SiN$_x$\footnotemark[1]       & 2000                     & LPCVD              & CF$_4$/Ar/O$_2$ ICP-RIE  \\
Nb wiring 1\textsuperscript{1,2}             & 200                      & sputter            & SF$_6$/O$_2$ ICP-RIE     \\
AlMn TES\footnotemark[1]                       & 400                      & sputter            & Transene Al Etch Type A \\
Ti/PdAu/Ti shunt resistor\footnotemark[2]      & 10/340/10                & sputter            & liftoff                  \\
SiN$_x$ insulator 1\textsuperscript{1,2}          & 350                      & PECVD              & CF$_4$/Ar/O$_2$ ICP-RIE  \\
Nb wiring 2\textsuperscript{1,2}             & 400                      & sputter            & SF$_6$/Ar ICP-RIE        \\
Ti/Au resistor\footnotemark[1]                 & 5/100                    & e-beam evaporate   & liftoff                  \\
SiN$_x$ insulator 2\textsuperscript{1,2}          & 400                      & PECVD              & CF$_4$/Ar/O$_2$ ICP-RIE  \\
PdAu heat capacity\footnotemark[1]             & varies                   & sputter            & liftoff                  \\
SiN$_x$/SiO$_x$ frontside punch\footnotemark[1]& -                        & -                  & CF$_4$/Ar/O$_2$ ICP-RIE  \\
SiN$_x$/SiO$_x$ backside\textsuperscript{1,2}     & -                        & -                  & CF$_4$/Ar/O$_2$ ICP-RIE  \\
Ti/Au backside metal\footnotemark[1]           & 5/400                    & e-beam evaporate   & liftoff                  \\
Si release\textsuperscript{1,2}              & -                        & -                  & SF$_6$ + C$_4$F$_8$ DRIE \\
Membrane release\footnotemark[1]               & -                        & -                  & -                        \\
\botrule
\end{tabular}
\footnotetext[1]{Fabrication steps used for detector arrays.}
\footnotetext[2]{Fabrication steps used for DC/RF routing wafers.}
\end{table}

The detector array fabrication process follows Ref. \cite{duff-advact}, which describes the rhomboidal detector array layout of repeating pixels and dense wiring buses patterned with stepped lithography in 11 separate steps. Each SO MF and UHF array has 430 optical pixels with 4 polarization sensitive TESs per pixel in 2 frequency bands, for a total of 1,720 superconducting detectors. The arrays are fabricated on double-side-polished (DSP), 500-$\mu$m-thick, 150~mm silicon (Si) wafers coated with 450~nm thermal silicon oxide (SiO$_x$) and 2~$\mu$m low-pressure chemical vapor deposited (LPCVD) super-low-stress silicon nitride (SiN$_x$), both deposited at a commercial vendor \cite{roguevalley}. These dielectric layers are used to define the thermal link between the isolated bolometer island, which houses the TES, and the thermal bath of the thick Si substrate. The monolithic 1,720-channel DC/RF routing wafers are fabricated on DSP, 500-$\mu$m-thick, 150~mm Si wafers coated with 150~nm thermal SiO$_x$ and utilize many of the same materials and processes as the detector arrays, but rely on a direct-write photolithography process to pattern the 6 layers of non-repeating circuitry \cite{healy_umm}. 

\begin{figure}[t]%
\centering
\includegraphics[width=0.99\textwidth]{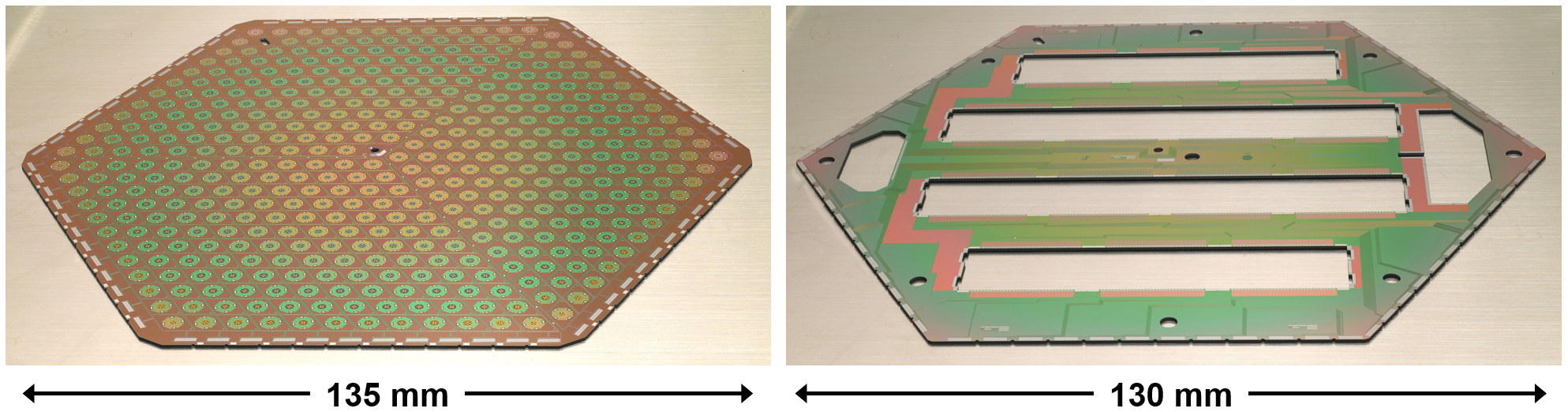}
\caption{SO UHF detector array (left) and SO DC/RF routing wafer (right). The large areas of removed Si in the DC/RF routing wafer are designed for drop-in placement of the multiplexing chips.}\label{fig:stack}
\end{figure}

Figure~\ref{fig:fab} shows a series of micrographs from a 90~GHz TES bolometer at each layer in the fabrication process. The frontside and backside processing steps are summarized in Table~\ref{tab:fab}. A SO UHF detector array and DC/RF routing wafer are shown in Fig.~\ref{fig:stack}.

\section{Metrology and Process Monitoring}\label{sec:metrology}
Detector characteristics such as TES normal resistance $R_{\mathrm{n}}$, superconducting critical temperature $T_{\mathrm{c}}$, saturation power $P_{\mathrm{sat}}$, and time constant $\tau_{\mathrm{eff}}$ are controlled with well-understood device design and fabrication processes. CMB instrument performance relies on detector arrays operating as designed. For example, controlling $T_{\mathrm{c}}$ is important, as detector noise equivalent power scales as $T_{\mathrm{c}}$$^{\mathrm{(n+1)/2}}$, where n $\sim{~}$3 \cite{irwinhilton}.

To ensure a fabrication process that results in repeatable high-yield wafers, we have adopted a 3-tier metrology approach. This allows for calculated decisions regarding which processes to monitor and measurements to collect during and after fabrication. It balances a high-yield repeatable fabrication process with delivery schedules, personnel skillsets, and equipment demands. 
Implementing this production fabrication framework has resulted in 88$\%$ (72$\%$) of all detector wafers (DC/RF routing wafers) meeting nominal criteria to be integrated into a module. Detectors must have $>95\%$ channel yield (room temperature electrical continuity, mechanical membrane yield, microscope inspection of channel defects) and $T_{\mathrm{c}}$ in the targeted range; DC/RF routing wafers must have $>90\%$ channel yield and shunt resistance in the targeted range.

\begin{figure}[t]%
\centering
\includegraphics[width=0.99\textwidth]{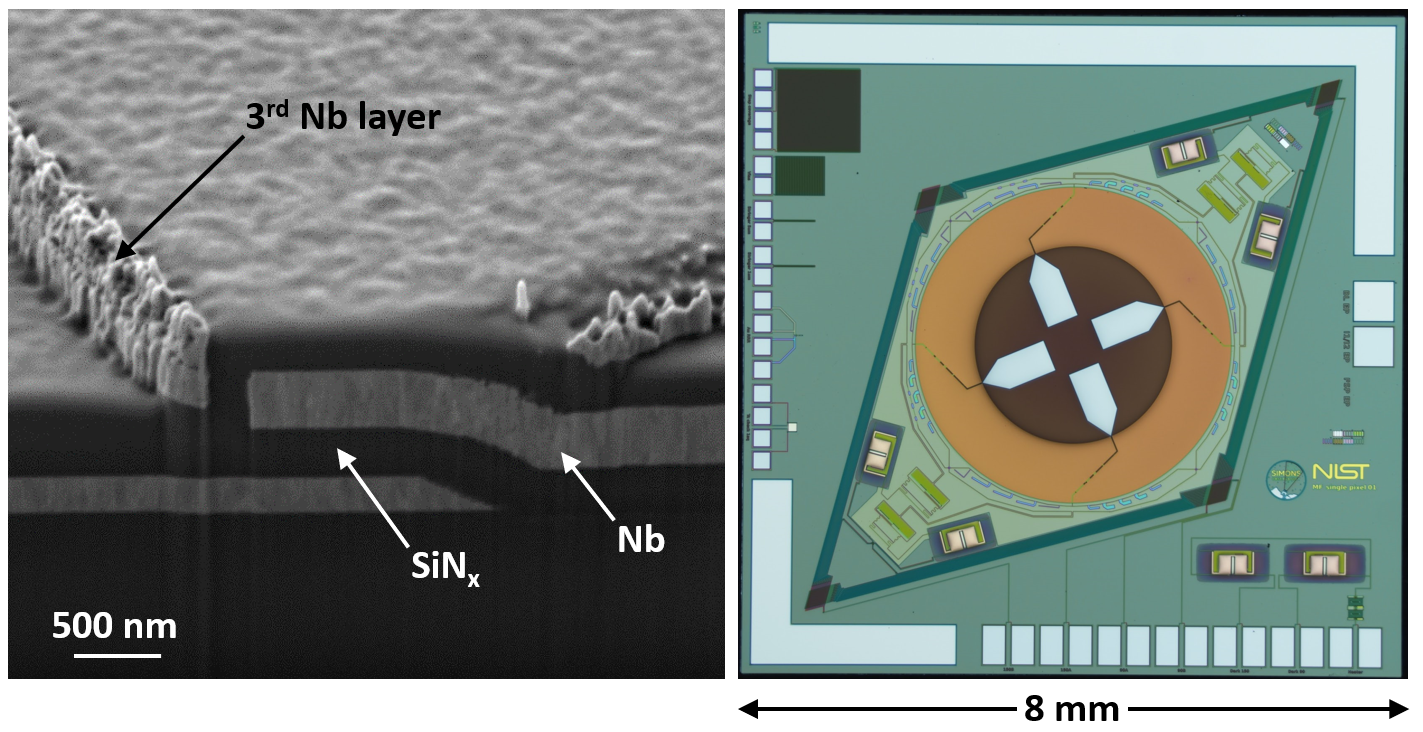}
\caption{Top tier metrology. (Left) A multi-layer cross-section FIB-SEM image shows Nb (light) and SiN$_x$ (dark) layer thickness and etch profile. This micrograph prompted the development of a simplified and more robust process to eliminate the third Nb wiring layer, seen here as an undesired remnant filament on the top surface which may cause electrical shorts. (Right) SO MF single test pixel fabricated to verify prototype design.
}\label{fig:tier1}
\end{figure}

The top tier of metrology includes process development tasks and pre-production prototyping, which validate single- and multi-step process quality and compatibility. 
These tasks occur when process or design changes are required. High-resolution scanning-electron microscope (SEM) imaging, with and without focused-ion beam (FIB) milling of single- and multi-layer cross-sections, allows for verification of etch processes, layer conformality, and visual detection of any defects that may compromise the device operation. Fabrication of single test pixels allows for pre-production rapid prototyping with dark and optical characterization of $T_{\mathrm{c}}$, $P_{\mathrm{sat}}$, on-chip passband filters, optical efficiency, and polarization response to feed back to detector array design \cite{walker-so, hubmayr-lb}. Measurements that match simulation confirm the design and the quality of the overall fabrication process. Examples of top tier metrology tasks are shown in Fig.~\ref{fig:tier1}.

\begin{figure}[b]%
\centering
\includegraphics[width=\textwidth]{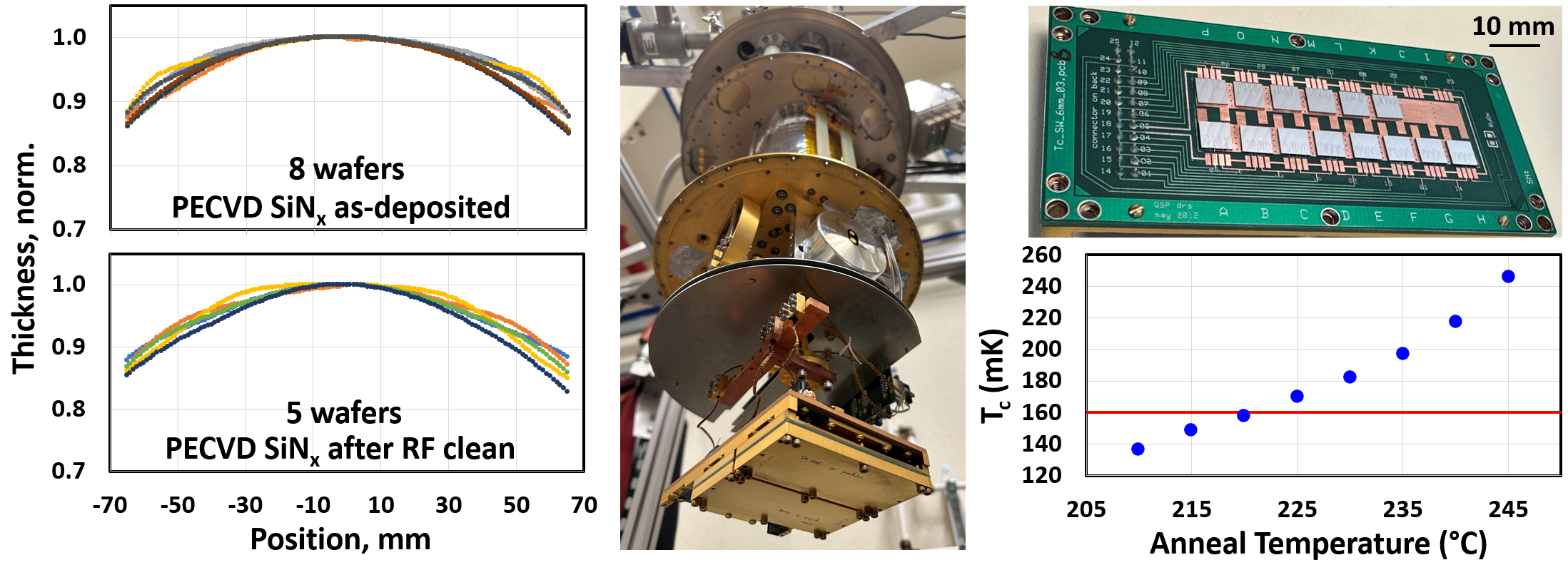}
\caption{Middle tier metrology. (Left) SiN$_x$ thickness uniformity as-deposited (8 separate wafers) and after in-situ Nb deposition RF clean (5 separate wafers), showing that 
radial uniformity is maintained at each step and that the fabrication process is highly repeatable. (Middle) Photograph of rapid turnaround cryostat used for in-process $T_{\mathrm{c}}$ measurements. (Right) Custom 4-wire PCB for superconducting film and device characterization and plot of Al with 0.2 atomic percent Mn anneal temperature vs. $T_{\mathrm{c}}$ for SO MF detector batch, with the red line indicating the target $T_{\mathrm{c}}$.}\label{fig:tier2}
\end{figure}

\begin{figure}[t]%
\centering
\includegraphics[width=\textwidth]{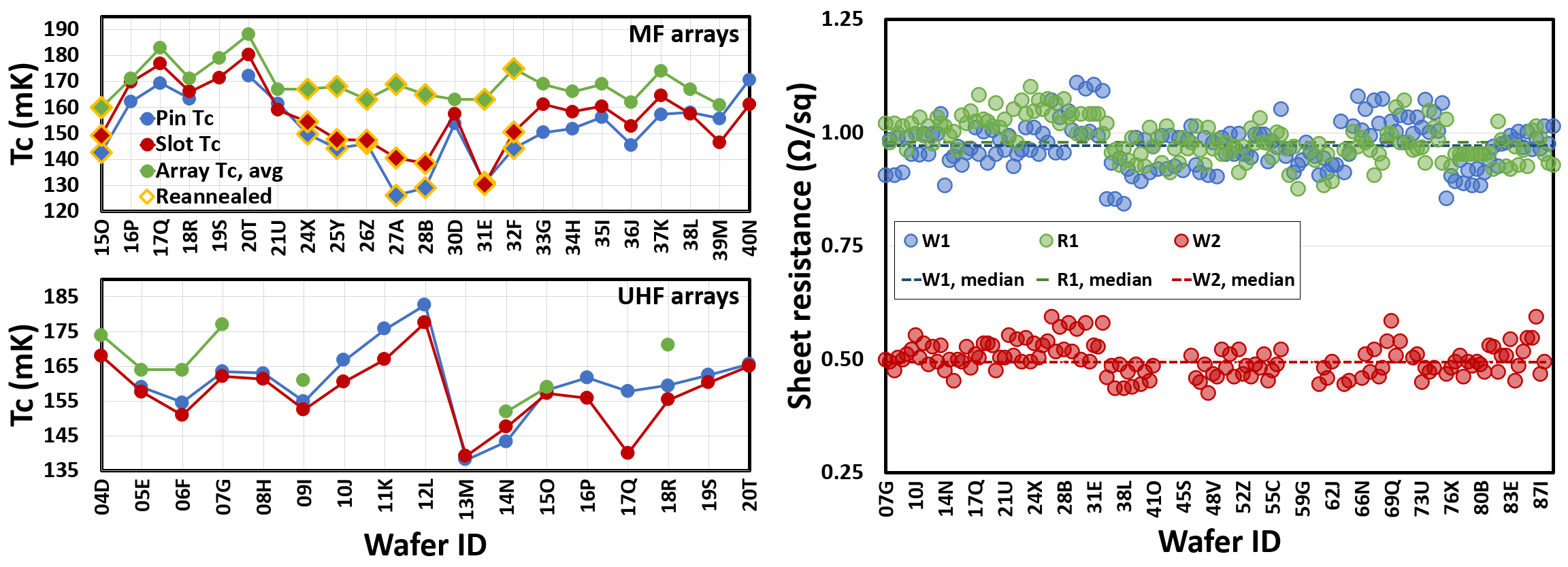}
\caption{Bottom tier metrology. (Left) Measured $T_{\mathrm{c}}$ of the center (pin), edge (slot), and average for each SO MF and UHF detector array, plotted in sequential order. Some SO MF detector arrays have been re-annealed post-fabrication to raise the $T_{\mathrm{c}}$ of the AlMn TES to the target value, 160~mK ${\mathrm{\pm}}$ 10~mK. (Right) Extracted values of sheet resistance from room temperature measurements of cross-bridge test structures on the DC/RF routing wafers \cite{crossbridge}. The dashed lines are the median value. 
}\label{fig:tier3}
\end{figure}

The middle tier consists of measurements that occur with every batch. CMB instruments rely on high detector efficiency, which can be achieved with low-loss superconducting microstrip wiring. As in \cite{chang-loss}, the loss of SiN$_x$ and Nb thin films at microwave and millimeter-wave frequencies is measured and correlated to detector optical efficiency. Additionally, it is critical to monitor thin film deposition rates, etch rates, and thickness uniformity across the wafer, as in Fig.~\ref{fig:tier2}. 
Characterization of the SiN$_x$ thickness across the wafer at each step in the process allows the fabricated device to match the design, which directly impacts the passband placement, e.g.
As mentioned, achieving a target $T_{\mathrm{c}}$ is essential and each detector array batch includes a separate witness wafer to calibrate the AlMn process by measuring $T_{\mathrm{c}}$ as a function of annealing temperature \cite{li-advact}. Since the AlMn $T_{\mathrm{c}}$ can be tuned with annealing, this gives a per batch prescription for achieving the targeted $T_{\mathrm{c}}$.

The SO production fabrication utilizes a detailed fabrication process sheet, where every essential processing detail is documented. Each batch of detector arrays and DC/RF routing wafers includes a fabrication summary document, providing critical information to collaborators like room temperature electrical continuity, mechanical membrane yield, and results from microscope inspection for channel defects.

The bottom tier metrology is critical on a per wafer or per layer basis, with examples as in Fig.~\ref{fig:tier3}. 
Measurements for each wafer include thickness of thermal SiO$_x$ and LPCVD SiN$_x$ as delivered from vendor; critical layer (AlMn, Nb, PdAu) sheet resistance; room temperature probing for filament shorts, Nb residue, via continuity, crossover isolation, and wiring residue; full wafer semi-automated room temperature electrical continuity probing; and $T_{\mathrm{c}}$ measurements from center and edge of the detector array. At the per layer level, metrology steps include thin film stress, profilometry, ellipsometry, and manual optical microscope inspection.

\section{Volume Production}\label{sec:production}
The NIST cleanroom, much like many other facilities fabricating low temperature detectors, is a shared R\&D facility and is not designed for high levels of wafer throughput. 
Scaling up to higher volumes necessitated developing a fabrication framework that balances time versus risk/reward for both the implementation of metrology steps and for process changes and improvements. 
The serial nature and maximum capacity of many NIST cleanroom tools was taken into account by optimizing the batch size to four wafers. 
Carrying multiple active in-fab batches in parallel and utilizing collaborative tool scheduling allows for maximum flexibility in optimizing tool usage and personnel time. 

Implementation of these efficiencies in the NIST cleanroom has allowed for high-yield production of complicated superconducting devices at the scale required for SO. Throughout the entirety of the SO production fabrication endeavor, the average realized production fabrication time for detector arrays and DC/RF routing wafers was 9 weeks and 6 weeks per batch, respectively. 

\section{Ongoing Improvements}\label{sec:dualtes}
\begin{figure}[t]%
\centering
\includegraphics[width=\textwidth]{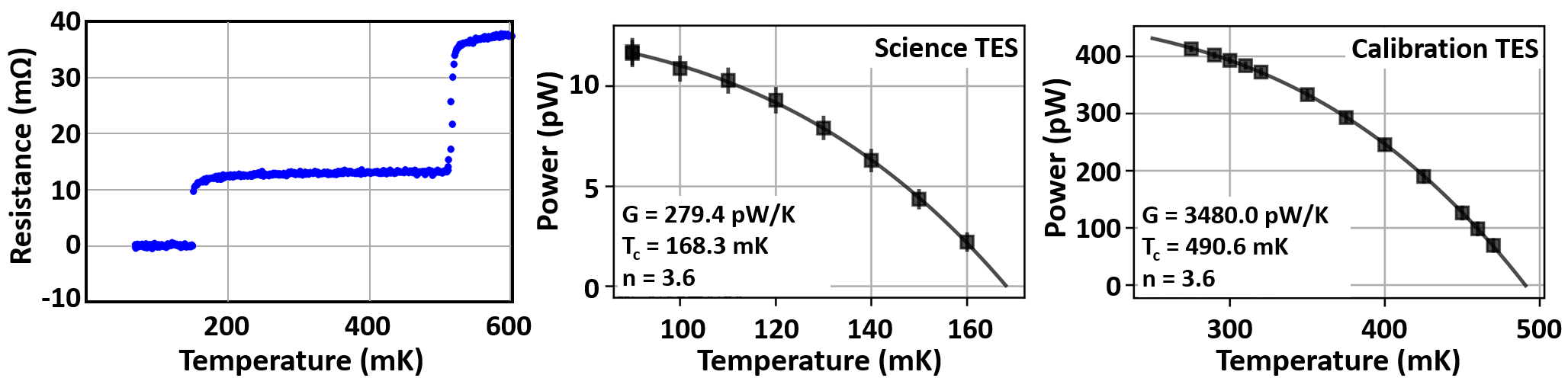}
\caption{(Left) Resistance versus temperature plot showing two superconducting transitions, one for the science TES at 150~mK and one for the calibration TES at 517~mK. (Middle) Power sweep for the science TES at a range of bath temperatures. (Right) Power sweep for the calibration TES at a range of bath temperatures.}\label{fig:dualtes}
\end{figure}

One benefit of a robust fabrication process is the capacity to add complexity in one or more steps. 
Experience with AlMn TESs has allowed for development of a second in-series AlMn TES with a higher $T_{\mathrm{c}}$ for increased dynamic range by adjusting the Mn concentration in Al. This enables more straightforward characterization of optical properties with standard in-lab loading conditions following \cite{jpl-dualtes}, but optimized for 100~mK refrigeration which has become standard for current and future CMB experiments baselining this technical advancement \cite{cmbs4a,taurus}.
Prototype single test pixels implementing the dual-AlMn TES process have been fabricated and characterized. The dual-transition and projected $P_{\mathrm{sat}}$ for both TESs are shown in Fig.~\ref{fig:dualtes}. 

\section{Conclusions}\label{sec:conclusions}
We have presented a robust fabrication framework at NIST which has resulted in successful delivery of more than 40 detector arrays and more than 50 DC/RF routing wafers, completing the first phase of SO. These wafers are high-yield and have high intra- and inter-wafer uniformity, with $88\%$ of detector wafers 
meeting nominal criteria for integration into a detector module: a channel yield $>95\%$ and $T_{\mathrm{c}}$ in the targeted range \cite{dutcher_somf}. By implementing volume-production best practices and metrology and process monitoring for maximum process control and repeatability, we have demonstrated a production fabrication rate of 9 weeks and 6 weeks per batch of detector arrays and DC/RF routing wafers, respectively.

\bmhead{Acknowledgments}
This work was supported in part by the Simons Foundation (Award \#457687, B.K.).

\bibliography{sn-bibliography} 

\end{document}